\documentclass{zermatt}
\usepackage{graphicx}
\usepackage{natbib}
\usepackage[T1]{fontenc}
\usepackage[utf8]{inputenc}
\usepackage[english]{babel}
\newcommand{\CII}{[C\,{\sc ii}]}
\newcommand{\NII}{[N\,{\sc ii}]}
\newcommand{\OI}{[O\,{\sc i}]}
\newcommand{\OIII}{[O\,{\sc iii}]}

\def\thisvolume{this volume} 
\newcommand\kms{km\,s$^{-1}$}
\newcommand{\mic}{$\mu$m}
\newcommand{\hho}{H$_2$O}
\newcommand{\pow}[2]{#1$\times$10$^{#2}$}
\newcommand{\gtsim}{{_>\atop{^\sim}}}
\newcommand{\ltsim}{{_<\atop{^\sim}}}
\begin{document}
\title{Star \& planet formation: Upcoming opportunities in the space-based infrared} 
\runningtitle{Van der Tak: Space-based infrared opportunities for star \& planet formation}
\author{Floris van der Tak}
\address{SRON Netherlands Institute for Space Research, The Netherlands; \email{vdtak@sron.nl}}
\secondaddress{Kapteyn Astronomical Institute, University of Groningen, NL}
\begin{abstract}
While ALMA and JWST are revolutionizing our view of star and planet formation with their unprecedented sensitivity and resolution at submillimeter and near-IR wavelengths, many outstanding questions can only be answered with observations in the thermal (mid- and far-) infrared domain. 
Many of these questions require space-based observations, to achieve the necessary sensitivity and/or wavelength coverage.
In particular, how do interstellar clouds develop filamentary structures and dense cores? 
What are the masses and luminosities of objects at the earliest stages of star formation? 
What are the gas masses of planet-forming disks, and how do these disks disperse during planet formation? 
How is refractory and volatile material distributed within the disks, and how does this evolve with time? 
This article reviews how upcoming and planned balloon-borne and space-based telescopes for the mid- and far-infrared will address these questions, and outlines which further missions will be needed beyond 2030, when the ELTs will be in full operation.
\end{abstract}
\maketitle
\section{Introduction: Why the space-based infrared}

Observations in the thermal (mid- and far-) infrared, defined here as the 3--300\,\mic\ wavelength range, are uniquely able to probe the cold obscured Universe. 
One area of astrophysics requiring such observations is the study of the evolution of galaxies from the formation of the first stars (`cosmic dawn', $z$=6--20) via the peak epoch of cosmic star formation (`cosmic noon', $z$=2--3) into the present-day galaxy population (Aravena, \thisvolume). 
Key questions include the universality and physical drivers of galaxy scaling relations such as the 'star formation main sequence' \citep[e.g.,][]{pearson2018}, and the origin of dust and heavy elements in galaxies (Henning, \thisvolume).
Atomic fine structure lines in the mid- and far- infrared are key diagnostics of ISM conditions at redshifts out to $\approx$6 \citep{andres2021,andres2022}.

The second area of astrophysics where thermal infrared observations play a large role is the physics of galaxies, in particular the so-called baryon cycle \citep{vdtak2018}.
Key questions in this area are how the star formation process depends on galaxy properties, and the role of feedback from stars and AGN \citep{saintonge2022}.
Spectroscopic maps in the mid- and far-infrared are crucial to address these questions, both of atomic (\CII, \NII, \OI, \OIII, ...) and of molecular (HD, OH, \hho, high-$J$ CO, ...) tracers, as outlined by Madden (\thisvolume) and Rubio (\thisvolume).

The third area is the formation of stars and planets, which is the topic of this conference and the focus of this article. 
In the area of star formation, ALMA has allowed great progress, in particular for the structure of accretion disks \citep[e.g.,][]{cesaroni2017,maud2019}, protostellar multiplicity \citep[e.g.,][]{tobin2022}, and the chemical structure of pre-stellar cores \citep[e.g.,][]{caselli2022}.
Similarly, breakthroughs are expected from JWST, if its first results are any guide \citep[e.g.,][]{yang2022}.
However, most ALMA results so far are on individual sources or small samples, which has the risk of creating a biased view. 
Larger surveys to build up statistics and remove selection biases are planned, in particular ATOMS \citep{liu2020}, ALMA-IMF (Stutz, \thisvolume) and ALMA-GAL (Fuller, \thisvolume).
Furthermore, several outstanding issues in star formation require far-infrared observations to be resolved, in particular the very early stages.
Witnessing the formation of molecular clouds out of atomic gas requires wide-field velocity-resolved maps of the fine structure lines of \CII, \NII, and \OI, as pioneered with SOFIA \citep{pabst2019,kavak2022} attempted with the STO-2 balloon mission \citep{seo2019} and planned with its successor GUSTO \citep{walker2022}.
Herschel surveys \citep{molinari2016,elia2021} have shown that these clouds develop a filamentary structure with hubs and branches, but unraveling the mechanism of their formation also requires wide-field maps of the same fine structure lines at $\sim$\kms\ spectral resolution \citep{hacar2022}.
Magnetic fields may also play a role in creating these structures, and polarimetric maps of arcmin-sized fields are needed to find out. 
SOFIA has pioneered this field \citep{stephens2022}, ground-based single dishes (e.g., CCAT) and balloon-borne missions (e.g., BlastPol) may take initial steps \citep{pattle2022}, but space-based observations are needed to fully address this question \citep{andre2019}. 
Finally, studying the collapse of clumps inside filaments and the fragmentation of clumps into cores, which is crucial to understand the origin of the observed low star formation efficiency and Initial Mass Function shape, needs sub-arcsecond imaging around 100\,\mic, which requires an interferometer or a large dish in space.

The formation of planets is another area where ALMA has allowed great progress, both in terms of disk physical structure \citep{andrews2018,vdmarel2021} and in terms of disk chemistry \citep{oberg2021,brunken2022}.
Many JWST programs focus on disks, and progress is expected especially for inner disks ($\ltsim$10\,au), where planets are thought to form.
The two facilities complement each other: ALMA probes the outer and JWST the inner parts of disks.
However, while both telescopes excel in revealing the detailed structure of individual objects, they struggle to carry out surveys to build up statistics. 

Even with JWST and ALMA in operation, far-infrared observations are essential to address key questions about protoplanetary disks \citep{kamp2021}.
First among these is the gas mass, which is best probed by the HD molecule, through its $J$=1--0 and 2--1 transitions at 112 and 56\,\mic\ \citep{trapman2017}.
Spatially and/or spectrally resolved HD observations of disks are needed to understand how their gas masses evolve during the process of planet formation.
A full view of the mid- and far-IR spectra of disks is needed to understand the evolution of the \hho\ gas and ice abundances, and the distribution of minerals and ices within the disk; and to link their dust composition to that of asteroids.
See \citet{miotello2022} for a review of bulk disk properties and the physical environment in which planets form.

\section{Outstanding questions and instrumentation needs}

While the need for mid- and far-IR observations is clear, specific instrument requirements strongly depend on the science case at hand.
In particular, if high angular and/or spectral resolution is needed, ground-based telescopes are the natural choice, as such performance scales with physical size.
Large mirrors are able to push diffraction limits, and large gratings offer high resolving powers. 
A case in point is the METIS instrument for the e-ELT, which just passed FDR and is scheduled for first light in 2027 \citep{brandl2022}.

Space missions offer two advantages: superior sensitivity (due to the low thermal background), and broad wavelength coverage (by lack of atmospheric absorption).
The current state of the art are the Spitzer and Herschel missions, which respectively offered a small cold mirror (85\,cm, 5.5\,K) and a large warm mirror (3.5\,m, 80\,K). 
The natural next step is a large ($\gtsim$2\,m) cryogenic ($\ltsim$10\,K) mirror.
In the past decade, two such concepts were developed: SPICA in Europe/Japan \citep{roelfsema2018} and OST in the US \citep{meixner2019}, neither of which however made it to agency adoption.

All three world-leading space agencies have plans for infrared space missions, but their timescales and levels of maturity differ greatly.
In Japan, JAXA is considering the Grex-Plus mission, which would offer a 1.2\,m mirror cooled to 50\,K, a 1,400 arcmin$^2$ camera for the 2--10\,\mic\ range, and an $R$=30,000 spectrometer for the 12--18\,\mic\ band \citep{inoue2022}.
The concept derives from the SPICA/SMI instrument, and launch is foreseen for the early 2030s. 

In Europe, ESA is starting its Voyage 2050 program\footnote{\tt https://www.cosmos.esa.int/web/voyage-2050}.
No mid- or far-IR concepts have survived the ongoing selection of the M7 mission, with launch in the late 2030s\footnote{\tt https://www.cosmos.esa.int/web/call-for-missions-2021/update-on-the-f2-and-m7-mission-opportunity}.
The L5 mission is very likely to be in the infrared: either a near-IR successor to the Gaia astrometry mission, or a mid-IR mission for exoplanet characterization. 
However, neither concept covers the far-IR nor addresses the above science topics; and in any case, the launch of L5 is not foreseen until the 2040s.

Third, following the Decadal report, NASA has announced an opportunity for Probe-class missions, with a cost cap of 1\,B\$ excluding launch and GO program\footnote{\tt https://explorers.larc.nasa.gov/2023APPROBE/pdf\_files/NNH22ZDA008L.pdf}.
Up to 30\% of the cost may be externally paid (similar to e.g. HST and JWST), but given its current financial troubles, ESA is unlikely to participate.
Launch of the Probe should be in the early 2030s, which requires a high technology readiness level at the proposal stage (2023-2025).
The announcement specifies that the Probe should be either an X-ray or a far-IR mission.


The instrumentation needs of a far-IR space telescope strongly depend on the science case.
Galaxy evolution studies generally need maximum sensitivity ($\sim$10$^{-19}$\,W\,m$^{-2}$) 
and survey speed (ideally of contiguous fields with complementary data), while modest spectral resolution ($\sim$200) is enough to measure line intensities.
In contrast, most studies of the ISM of local galaxies need a reasonably high mapping speed ($\sim$arcmin/hr) and spectral resolution ($\sim$3000), while moderate angular resolution ($\sim$10$''$) tends to be sufficient.
Infrared studies of Galactic star formation typically require wide ($\sim$degree) fields and $\sim$\kms\ spectral resolution, while today's sensitivity levels are adequate.
Planet formation studies  require high ($\sim$10$^{-19}$\,W\,m$^{-2}$) sensitivity and $\sim$\kms\ spectral resolution, while survey speed is irrelevant as known sources are observed.

Clearly, the disparity in requirements between the four main areas of mid/far-infrared astrophysics is too large to bridge by compromise.
Adding other areas where space-based infrared observations are important, in particular Solar system science and exoplanet characterization, only exacerbates the situation. 
Given finite budgets, future infrared missions therefore are unlikely to be general-purpose observatories like Spitzer and Herschel, but will instead be optimized to address selected scientific areas.

\section{NASA opportunities}

This section describes the four infrared concepts that are currently being developed for NASA's Probe-class opportunity. 
Note that the detailed specifications of the telescopes and instruments are likely to evolve as the concepts mature.
For a review of the underlying detector physics, see Staguhn (\thisvolume).

The PRobe Infrared Mission for Astrophysics (PRIMA; \citealt{bradford2022}) is a cryogenic 2\,m telescope which will be confusion limited at wavelengths $\gtsim$70\,\mic.
To optimally benefit from the cold mirror, the team plans to use Kinetic Inductance Detectors (KIDs) with an NEP of $\sim$10$^{-19}$\,W\,Hz$^{-1}$.
Such detectors have been demonstrated in the lab \citep{baselmans2022}, and space qualification is underway.
Two main instruments are foreseen for PRIMA: a 25--230\,\mic\ imager with polarimetric capabilities at the long-wavelength end; and a grating spectrometer giving $R$=60--250 over the 25--330\,\mic\ range.
In addition a high-resolution (Fourier transform or Fabry-P\'erot) spectrometer is planned, giving $R$=3000--5000 at a reduced sensitivity, mapping speed, and spectral coverage.

The science case for PRIMA is broad, but the main focus is on galaxy evolution. 
The high sensitivity, spectral capability, and high blind spectral survey speed are powerful tools to study the coevolution of star formation and black hole accretion in galaxies \citep{bisigello2021}, as well as the buildup of heavy elements and dust at high redshift. 
The imaging capability allows surveys of cosmological deep fields, with matching photometry at optical/UV and radio wavelengths.
For local galaxies, PRIMA is well suited for studies of ISM conditions and dust content.
For Galactic star formation, PRIMA is especially powerful to study protostellar accretion variability \citep{johnstone2022} and the origin of magnetized ISM filaments.
For planet formation, PRIMA's grating spectrometer is useful to study disk mineralogy, while gas masses can be measured with the high-resolution mode.

The FIRSST concept (Far-Infrared Spectroscopic Survey Telescope) is also a cryogenic $\sim$2\,m telescope, with a raw sensitivity similar to PRIMA, i.e., $\sim$100$\times$ better than Herschel/PACS. 
Its main instrument is a direct-detection imaging spectrometer, which in its broad-band mode covers the 30--270\,\mic\ range at $R$=200, suitable to measure fine structure line intensities in targeted surveys of high-redshift galaxies.
The medium-resolution mode offers $R\sim$2000 which is suitable to study feeding and feedback processes in local galaxies, as pioneered in OH lines by \citet{sturm2011}.
The high-resolution mode uses VIPAs (virtual phased arrays) to provide $R$=10$^5$ over a 10\% bandwidth, centered on lines of interest such as HD, \CII, \OI, \hho, and HDO. 
This mode is especally useful to measure the gas masses of protoplanetary disks (in HD), and to study their dispersal (in \OI).  
The VIPAs are the most unique part of FIRSST, but they still require demonstration in the lab and space qualification.
In addition, FIRSST is planned to have a heterodyne spectrometer, offering $R$=\pow{3}{7}, with \hho\ and HDO chemistry as its main science goals.
The added value of this instrument seems somewhat limited, especially compared to Herschel/HIFI, which had higher angular resolution.
The sensitivity of heterodyne systems is quantum-limited, so that they do not fully benefit from the capabilities of a cryogenic mirror.

The Single Aperture Large Telescope for Universe Studies (SALTUS; \citealt{kim2022}) concept uses an inflatable 20-m non-cryogenic (45\,K) telescope with a deployable boom/torus structure.
This design builds on heritage from the 14-m Inflatable Aperture Experiment (IAE) in 1996 and the successful deployment of the 6.5-m JWST in early 2022 (see \citealt{quach2021}).
The sensitivity is similar to PRIMA and FIRSST, i.e. 10$\times$ better than SPICA, but the angular resolution is 1.4$''$, i.e. 10$\times$ better than PRIMA and FIRSST, and close to the optical and radio views of galaxies, rather than being an order of magnitude worse typically.
Three instruments are planned: (i) a 4-band KID-based grating spectrometer covering the 30--300\,\mic\ range at $R$=300, which is suitable to study the origin of dust and heavy elements in galaxies, as well as interstellar ices and minerals, and dust in planet-forming disks;
(ii) a set of tunable 8-pixel Hot Electron Bolometer arrays, covering selected bands in the THz window at $R$=10$^5$--10$^6$, suitable to study \hho\ chemistry in regions of star and planet formation;
(iii) a set of tunable SIS arrays covering the 520--650\,\mic\ band for ISM spectroscopy and the 870--1300\,\mic\ band to provide long baselines to the Event Horizon Telescope \citep{eht2019}.
The improved angular resolution would increase the EHT target list from 2 to $\approx$100 sources.
Optimizing the orbit of SALTUS will be a challenge: while the EHT connection requires proximity to Earth, the other science goals are better served further away, where the thermal environment is more benign.
While SALTUS is not suitable for blind surveys, its strength will be to observe large samples of individual (point) sources, e.g. to build up statistics for galaxy evolution studies.
With careful sample construction, the impact of selection bias can be limited.

The SPICE concept (SPace Interferometer for Cosmic Evolution)\footnote{\tt https://asd.gsfc.nasa.gov/spice/} is an interferometer with two connected elements on a maximum baseline of 36\,m, providing an angular resolution of 0.3$''$ at 100\,\mic.
The 1\,m dishes are cooled to 4\,K, giving a line sensitivity of $\sim$\pow{4}{-19}\,W\,m$^{-2}$. 
Spectral coverage is 25--400\,\mic\ at $R$=3000. 
Such a telescope is especially powerful for planet formation, as it can spatially resolve gas masses (in HD) as well as \hho\ gas and ice abundances for $\sim$100 disks. 
The concept is also of interest for galaxy evolution studies, as it measures atomic fine structure line emission as well as far-IR dust continuum at a resolution matched to optical and radio surveys.
The drawback is the poor instantaneous \textit{uv} coverage, so that a fully sampled image of the 1$'$ field takes $\approx$24\,hours to make. 
Deep extragalactic fields will require several passes, which limits the statistics that can be built up during the 3-year lifetime.
Observations of brighter sources such as QSOs, lensed galaxies, and local AGN will be much more efficient, as are spectra of T~Tauri stars and protoplanetary disks.
In continuum, the SPICE team is planning four-band multi-wavelength synthesis, enhancing its efficiency for e.g. signs of exoplanets in debris disks.
 

\section{Conclusions}

While the decommissioning of SOFIA is a setback, JWST is a success, and in $\sim$6 years, METIS will provide superb angular and spectral resolution in the ground-based 2.9--14\,\mic\ windows.
For planet formation, significant steps are expected for the physical structure and chemical evolution of protoplanetary disks, including signatures of protoplanets.
The METIS instrument will also be powerful for high-mass star formation, especially the nature of accretion flows and disks, and the formation of stellar clusters.

Beyond 2030, the future for mid- and far-infrared studies of star and planet formation is less clear, although several ideas exist. 
If adopted, JAXA's Grex-Plus mission will sharpen our view on planet formation, in particular the structure, kinematics, and composition of inner disks.
Later in the 2030s, NASA's Probe mission will be powerful for star and planet formation, provided an infrared concept is selected.
For this field, SALTUS is the preferred option, as it provides high angular and spectral resolution.
The deep high-resolution spectra from FIRSST will be useful to measure disk masses and water abundances.
SPICE can provide high-resolution dust maps of protostellar clusters and ice maps of protoplanetary disks.
With PRIMA, the dust and ice mineralogy of disks can be studied, as well as magnetized interstellar filaments and protostellar accretion variability.

The Probe concepts differ in their telescope optics: PRIMA and FIRSST are made for mapping, while SALTUS and SPICE aim at point sources.
They also have different technological challenges: SALTUS the deployment, and the others the cooling of one large or several small mirrors.
In terms of detectors, KIDs and VIPAs need further development, while heterodyne systems are ready for space \citep[e.g.,][]{gan2021}.

Parallel to these developments, suborbital platforms will be valuable to carry out specific single-instrument science cases.
Stratospheric balloon missions such as ASTHROS (and, if adopted, POEMM) will also be useful for prototyping and space qualification of new instrument and detector technology.
Significant work is needed to turn these concepts into reality.
We conclude that the near- and mid-term future for the near- and mid-IR is bright, but that the far future is more uncertain, especially for the far infrared.

\bigskip

\textbf{Acknowledgements:} The author thanks Inga Kamp (Kapteyn), Peter Roelfsema and Lingyu Wang (SRON), and Lee Mundy (Maryland) for helpful comments.

\bibliographystyle{aa}
\bibliography{zermatt}
\end{document}